\documentclass[sigconf]{acmart}

\usepackage{wasysym}

\newcommand{\etal}{et al.}
\newcommand{\etals}{et al.'s}

\newcommand{\ie}{{i.e.}}
\newcommand{\eg}{{e.g.}}

\newcommand{\cf}{{cf.}}
\newcommand{\ala}{{\`a la}}

\newcommand{\secref}[1]{\hyperref[#1]{Sec.~\ref*{#1}}}
\newcommand{\appendixref}[1]{\hyperref[#1]{Appendix~\ref*{#1}}}
\newcommand{\figref}[1]{\hyperref[#1]{Fig.~\ref*{#1}}}
\newcommand{\eqnref}[1]{\hyperref[#1]{Eqn.~\ref*{#1}}}
\newcommand{\tabref}[1]{\hyperref[#1]{Table ~\ref*{#1}}}

\newcommand{\blackdiamond}{\mathord{\sbox0{$\diamond$}\resizebox{!}{1.125\ht0}{\raisebox{\depth}{\rotatebox[origin=c]{45}{$\blacksquare$}}}}}

\def\subsubsec#1
{\subsubsection{#1}}

\definecolor{linkColor}{HTML}{257E98}

\definecolor{skiGreen}{HTML}{4bd56e}

\AtBeginDocument{}

\copyrightyear{2025} 
\acmYear{2025} 
\setcopyright{rightsretained}
\acmConference[CHI EA '25]{Extended Abstracts of the CHI Conference on Human Factors in Computing Systems}{April 26-May 1, 2025}{Yokohama, Japan}
\acmBooktitle{Extended Abstracts of the CHI Conference on Human Factors in Computing Systems (CHI EA '25), April 26-May 1, 2025, Yokohama, Japan}\acmDOI{10.1145/3706599.3716230}
\acmISBN{979-8-4007-1395-8/25/04}

\begin{document}

\title[Linting is People!]{Linting is People! Exploring the Potential of Human Computation as a Sociotechnical Linter of Data Visualizations}

\author{Anamaria Crisan}
\affiliation{\institution{University of Waterloo}
  \city{Waterloo}
  \country{Canada}}
\email{ana.crisan@uwaterloo.ca}

\author{Andrew McNutt}
\affiliation{\institution{University of Utah}
  \city{Salt Lake City}
  \country{Utah}}
\email{andrew.mcnutt@utah.edu}

\begin{abstract}
  Traditionally, linters are code analysis tools that help developers by flagging potential issues from syntax and logic errors to enforcing syntactical and stylistic conventions.
  Recently, linting has been taken as an interface metaphor, allowing it to be extended to more complex inputs, such as visualizations, which demand a broader perspective and alternative approach to evaluation.
  We explore a further extended consideration of linting inputs, and modes of evaluation, across the puritanical, neutral, and rebellious dimensions.
  We specifically investigate the potential for leveraging human computation in linting operations through Community Notes---crowd-sourced contextual text snippets aimed at checking and critiquing potentially accurate or misleading content on social media.
  We demonstrate that human-powered assessments not only identify misleading or error-prone visualizations but that integrating human computation enhances traditional linting by offering social insights.
  As is required these days, we consider the implications of building linters powered by Artificial Intelligence.
\end{abstract}



\received{20 February 2007}
\received[revised]{12 March 2009}
\received[accepted]{5 June 2009}

\maketitle

\section{Introduction}

Linters are a means of collaboratively establishing correctness over a digitally analyzable object in a standardized and enforced manner.
Traditionally, linters have been used exclusively in the domain of code, acting as a sort of digital spell checker for errors that can be identified through static analysis.
For instance, identifying stylistic issues (like missing semicolons), functional issues (such as undefined variables), and domain-specific issues (such as certain operations being slow~\cite{cssLint}).
Recent work has explored generalizing the idea of a linter out of code and into other domains, such as visualization~\cite{chen2021vizlinter}, spreadsheets~\cite{barowy2018excelint}, and color palettes~\cite{mcnutt2024mixing}.

\begin{figure}[t]
  \centering
\includegraphics[width=\linewidth]{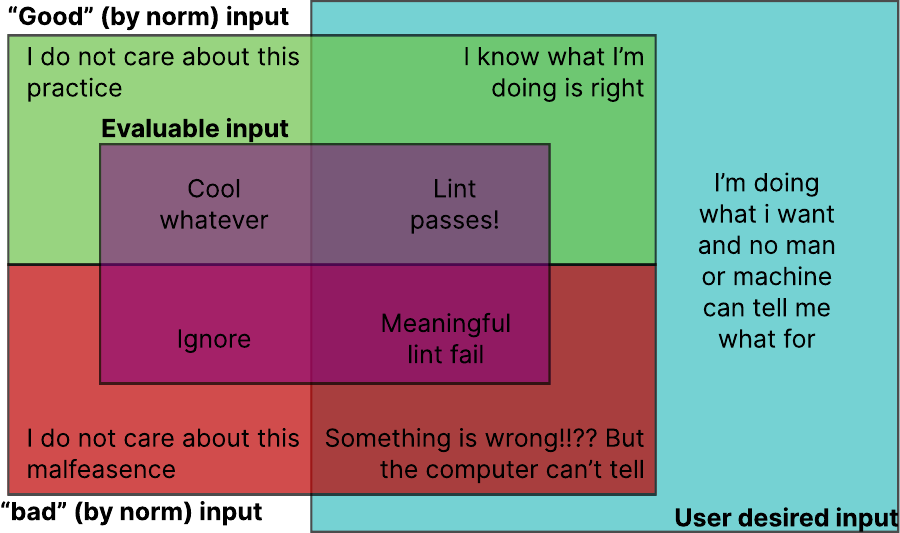}
  \Description{An euler diagram (generalized venn diagram) describing several different intersecting regions. There are four box label "'Good' (by norm) input", "Evaluable Input", "'bad' (by norm) input", and "User desired input". The boxes are arranged such that there are 9 intersecting regions. }
  \caption{A characterization of the input space for linting as Euler diagram. Linters tend to exist in domains where the input space is expressive enough for there to be types of input that are not correct by default. These rich domains are common, for instance, language, text, visualization, film, and many other mediums support such expansive input.
    Within these spaces linters can only reason about a subset of possible input (which is itself only a subset of the space for which there are community norms), which is possibly intersecting the space of input the user wishes to make. }
  \label{fig:lint-euler}
\end{figure}

Crucially, linters are a useful tool when the expression domain has been widened enough that mistakes can be of consequence (\cf{} \autoref{fig:lint-euler}).
There are, naturally, two areas in which this might apply: how an idea was expressed (\eg{} the sentence ``is squeak doodads when altercation'' is non-syntactically valid) and the content of the expression (\eg{} the sentence ``colorless green ideas sleep furiously'' is famously syntactically valid but semantically meaningless~\cite{wikiColorless_green_ideas_sleep_furiously}).
In prior work~\cite{mcnutt2024mixing} we argued that, as analysis tools improve, the most useful approach for linters is to analyze the latter of these. That is, to provide useful operationalization of knowledge about the execution domain rather than the input plane.

Yet! It is often not clear what it means for something to be right.
Notions of correctness, particularly in sociotechnical tools (like visualizations), are ambiguous in general, and community- and usage-specific in particular.
For instance, the rainbow color map is commonly argued~\cite{borland2007rainbow} in the visualization community as being something to be avoided (as it can cause hallucinations about the data via steep perceptual cliffs and so on).
Yet! In some communities, particularly in weather~\cite{ware2023rainbow, ware2013designing}, using standard color ramps of this format is nearly mandatory.
We suggest that---in interacting with scientists working in weather-adjacent areas---this is in part because using it indicates that you are a member of the in-group that you are trying to communicate with---but also because the people in that community are trained and familiar with reading information presented in that fashion.
Another example: the difficulty of ski runs in North America is delineated via a scale ({\color{skiGreen} \CIRCLE}-{\color{blue} $\blacksquare$}-{\color{black}$\blackdiamond{}$}). While this is broadly understood (and because of the double encoding of shape and color is color vision deficiency friendly), it generally does not follow conventions in perceptually comprehensible sequential ordering. Evidently, cultural convention can win out over scientifically motivated usage!
This highlights the boundaries of scientifically motivated linters: cultural knowledge can be hard to capture in simple computer-based checking tools, highlighting that there may be a need to explore other means of running a linter.

In this work we do just that: we extend this idea of linters being socially positioned to cover, not just the usage of lint, but their execution as well.  We consider the potential for leveraging human computation as a means of linting data visualizations to surface instances of valid visualizations (\eg{} adhering to established design guidelines), but also to assess whether it is meaningful to the intended audience. Human computation leverages groups of people to perform computational tasks, for example, data labeling, but has not been previously explored in the context of linting.

To do so we reflect on Community Notes (n\'ee Birdwatch)~\cite{twitterNotes}, which is a crowd-sourced moderation program that allows people to add context about potentially misleading content appearing on social media.
While Community Notes serve a variety of purposes,  we consider specific instances of their use to ``lint'' data visualizations (\autoref{fig:community-note-example}).
To explore the implications of linting as a metaphor, we reflect on its applications across the dimensions of its inputs and evaluative methods. Beginning with traditional code linters, which we consider to represent puritanical ideas of linting inputs and evaluative methods, we extend our exploration to more neutral and even rebellious perspectives.

\section{Related Work}

This work draws on linting (and related abstractions), as well as the idea of human computation.

\subsection{Linting and Similar Abstractions}

A key idea in this work is that linters are sociotechnical mediators of  community standards.
A good example of this can be seen in linters that operate over documentation.
The earliest of these that we are aware of, write-good~\cite{writeGood19Ford}, offers lint-style feedback on effective writing in markdown files (capturing issues like inappropriate use of passive voice or turns of phrase).
The Inclusive Language Plugin~\cite{elevntyInclsuiveLanguage} for the static site generator Eleventy automatically pushes users to use inclusive language (avoiding clauses like ``of course'' or ``just'').
HaTe Detector~\cite{winchester2023hate} guides documentation authors away from using hateful terms, such as master/slave instead of primary/replica.
Textlets~\cite{han2020textlets} includes functionality that supports automated checking of adherence to a small collection of invariants within the particular domain of legal writing.
The notion that linters are socially positioned objects is not new. For instance, the documentation of SQL Fluff~\cite{Cruickshank23SQLFluff} explicitly highlights that linters are a means to achieve communal adherence to a single style of authoring SQL queries.
We emphasize that by automating correctness, linters act as a form of policy enforcement.

Closely associated with the notion of linters, is the notion of standards. \citet{jackson2015standards} explore the meaning of standards and highlight how they can evolve over time and how standards need to be tuned to their local context (even when part of a larger project). A crucial part of standards is knowing when and how they can be broken appropriately.
We see this as being closely aligned with the interface philosophy of linters: they offer a clear signal of correctness from a particular perspective, but can be altered or ignored to fit a specific situation.

Moreover, a key facet of using linters as an interface metaphor is that, for a given domain, there is a collection of best practices or domain knowledge that needs to be operationalized (\ie{} a standard exists). This is a component of what differentiates them from a test suite: a linter is for a domain of work, whereas a test suite is for a project within a domain.
An additional assumption in comparing linters and test suites is that linters are quick while test suites are slow.
While this tension is worth exploring, in this work we forgo this component of this metaphor---instead focusing on the evaluatory and input aspects.

Within visualization there have been a number of different linting or lint-like systems.
In earlier work McNutt \etal{}~\cite{mcnutt2024mixing} survey these linters.
These include tools for checking chart usage based on a notion of best practices~\cite{mcnutt2018linting, zheng2019visualization, ChartLinter}, as well as more specific domains within visualization including exploratory modeling~\cite{kale2023evm}, choropleths~\cite{lei2023geolinter}, as well as the design of lint-style feedback~\cite{hopkins2020visualint}.
\citet{hull2023visgrader} develop a system for automatically grading visualizations created in an online course.
Connecting with documentation-centric linters, LitVis \cite{wood2018design} includes a notion of linting over the visualization design process, requiring the designer to provide documentation (via markdown comments) that follows particular patterns (such as a Socratic dialog or consideration of data feminist principles).
\citet{sultanum2024data} propose a notion of data guards, automated checks for fostering trust in data artifacts (such as visualizations). This view of trust as being a key mediating factor in the acceptance of data artifacts is closely related to our perspective on linters being devices that mediate community standards, in that both situate technical perspectives on manipulation of sociotechnical behaviors.
\citet{kosara24Boring} criticizes the idea of linting visualizations, suggesting that being critical without offering solutions is not a helpful usage paradigm (although some linters do exactly that).

Some approaches offer linter-style feedback but with a different pose relative to the development or analysis cycle.
\citet{lisnicvisualization} develop guard rails for preventing cherry-picking data.
\citet{narechania2021lumos} construct a data exploration system that highlights areas in data space not yet considered, surfacing potential biases.

Also closely related to linters is the question of teaching code quality.  \citet{keuning2023systematic} survey code quality in education, finding that some systems use linter or lint-like systems.
For instance, \citet{mcnutt2023study} conduct a study of creative coding editing tools with high school and college students, including a code-based linter. They find that students value linters as a means to internalize feedback.
We emphasize that this type of automation can lead to automation biases---\emph{if, as a beginner, the machine is telling me this is wrong then it's probably wrong, right?} This can be beneficial if the goal is to teach, but can be detrimental when the goal is a wrought artifact.
Automation bias, of which this is an example, is a problem that plagues many linters~\cite{chen2021vizlinter, lei2023geolinter}, not just those related to teaching.

Outside of visualization and text analysis, linters have been applied in a few different areas.
Wang \etals{}~\cite{wang2024farsight} Farsight provide lint-style warnings when a prompt for an AI system might lead to harm.
Data Linter~\cite{hynes2017data} checks to ensure that machine learning data does not have determinant components, such as NaNs.
ExceLint~\cite{barowy2018excelint} uses entropic analysis to identify likely errors in spreadsheets.
There are many areas and venues in which lint-style feedback can be applied left to explore. In this work we explore them in Community Notes.

\subsection{Human Computation}\label{sec:rw_hc}
Human computation is a widely used strategy that leverages collective human expertise and capabilities to complete tasks that are difficult for an automated method (\eg{} AI agent) to perform~\cite{Quinn:HC_taxonomy:2011}. Modern approaches use crowd-sourcing platforms (\eg{} Mechanical Turk, Prolific) to distribute a micro task among many people and aggregate the results to formulate a final response~\cite{lazar:human_comp:2017}.
Examples of human computation include dataset annotation and labeling~\cite{snow:labelling:2008,Chang:Revolt:2017}, document editing~\cite{Bernstein:Soylent:2010}, identifying accessibility issues~\cite{Hara:HC_accessibility:2013}, or even visual design critiques~\cite{Xu:Voyant:2014,Luther:CrowdCrit:2015}.
Response quality is an important factor, especially for complex tasks. Prior applications have shown that even non-expert workers can provide meaningful feedback to domain experts for complex tasks~\cite{Xu:Voyant:2014,Luther:CrowdCrit:2015}. However, there are a number of factors related to both the worker and the task itself~\cite{Allahbakhsh:CrowdIssues:2013}; to a reasonable extent these can be addressed through the design of applications using human computation.

While numerous visualization research studies have used crowd-sourcing for conducting experiments (\eg{} ~\cite{Heer:GraphicalPercention:2010,Borgo:Crowdsourcing:2018}), leveraging human computation in visualization-specific tasks has been explored relatively less.
For example, \citet{Willet:VisExplain:2012} leverage human computation to generate explanations for data visualizations. \citet{Wu:Layouts:2021} uses human computation to learn the effective layouts of charts. The application of human computation to linting tasks more generally, and to visualization linting specifically, has not been explored. Yet this may be an intriguing avenue, given the breadth of applications where human computation has been applied and the overlap these approaches have with broad goals of linting.

A key consideration for leveraging human computation is to design an environment and method to manage crowd workers and tasks~\cite{Allahbakhsh:CrowdIssues:2013}. While the majority of human computation approaches primarily leverage crowd-working platforms (such as Prolific or Amazon Mechanical Turk) to do so, other avenues exist.
For example, \citet{Moghadam:AutoStyle:2015} used a Massive Open Online Course (MOOC) platform to gather and analyze a corpus of code examples and develop a prototype that assists students by providing targeted feedback.
Even more prominently~\cite{ResearchMethodsHCI_Chapter14} CAPTCHA (Completely Automated Public Turing test to tell Computers and Humans Apart) uses human computation as biomarker, but the resulting datasets collected from tasks are used to train models.
More naturalistic scenarios for human computation, for example via social networks or other forms of human organization, have not been extensively explored or discussed.
One exception is LabintheWild~\cite{reinecke2015labinthewild} which mixes a notion of social challenge with study incentive to complete different human-computation driven tasks broadly for scientific purposes.
This rarity may be because it is neither intuitive nor natural to think of such organic interactions as a robust basis for reliable computational results. Yet, in the modern day, technology is closely woven into human interactions, whether they occur in person or online, and it is entirely possible for organic instances of even complex human computation to exist.

We consider a naturalistic application of human computation via social networks. We examine Community Notes and explore how they constitute both linting and human computation, setting aside (momentarily) questions of quality.

\begin{figure*}[t]
  \centering
  \begin{tabular}{p{0.15\linewidth}|p{0.25\linewidth}p{0.25\linewidth}p{0.25\linewidth}}
                                                   & \textbf{Evaluation Purist}                                                            & \textbf{Evaluation Neutral}               & \textbf{Evaluation Rebel}                      \\
                                                   & By static analysis                                                                    & By a computer                             & By whatever                                    \\\hline
    \textbf{Input Purist}\newline Input is code    & ESLint, gcc, VizLinter~\cite{chen2021vizlinter}                                       & Mirages Linter~\cite{mcnutt2020surfacing} & StackOverflow downvotes                        \\
    \textbf{Input Neutral} \newline Input is text  & HateDetector~\cite{winchester2023hate}, word good~\cite{writeGood19Ford}, spell check & Farsight~\cite{wang2024farsight}          & Community Notes (\secref{sec:community_notes}) \\
    \textbf{Input Rebel}\newline Input is whatever & ColorBuddy~\cite{mcnutt2024mixing}                                                    & Prof. in Art Sqool~\cite{artSqool}        & Architecture Planning Review                   \\
  \end{tabular}
  \Description{A table showing different systems. It has columns labeled "Evaluation Purist By static analysis", "Evaluation Neutral By a computer", and "Evaluation Rebel By Whatever". It has rows "Input purist input is code", "input neutral input is text", "input rebel input is whatever"}
  \caption{Different systems organized in an alignment chart format~\cite{kym_alignment_chart}. This framing forgoes nuances between categories---such as human-machine teaming which might exist between evaluation neutral and rebel---and leaves consideration of those interactions to future work. }
  \label{fig:alignment-chart}
\end{figure*}

\section{Abstracting Linting}

Here we offer a categorization of the linting design space, lensed through the way in which the lint is evaluated and the context in which it is used.
Following a common meme format~\cite{kym_alignment_chart}, we organize these axes along measures of adherence to orthodoxy  (ranging from purist to rebel).
On one axis we consider the notion that linters are something that uses static analysis (\ie{} code is not executed) to evaluate code (which is typical for code linters or compilers).
On the other, we consider the means by which the specific target of analysis.
We show this chart in \autoref{fig:alignment-chart}.

In prior work, \citet{mcnutt2024mixing} developed a linting ladder, describing a collection of four properties that broadly describe the behaviors expectable from a linter.
This model is the only model of linting usage as far as we are aware.
\textbf{Checkable} observes that a linter can evaluate whether or not a target has run afoul of the considered property.
\textbf{Customizable} is the ability to selectively deactivate lints, for instance by noting that a property does not matter in a given context.
\textbf{Blamable} is the affordance of specificity, highlighting the origin of the issue.
Last is \textbf{Fixable} or the propensity for automated repair of an identified issue.
For instance, consider the code linter ESLint's reaction to a missing semicolon.
It can identify that something is missing (checkable), it can be told to ignore that feedback because the style of the current project is to not use semicolons (customizable), where that error occurs (blamable), or even propose a fix like adding a semicolon (fixable).
There are other conceivable rungs in this ladder, such as counterfactuals, but these inscribe the behavior of most linters.
With this model in hand we now explore a number of our identified linters from \autoref{fig:alignment-chart}.

The upper quadrant of \autoref{fig:alignment-chart} is generally well explored by prior work~\cite{mcnutt2024mixing}, and so will spend most of our time with the more boundary-pushing parts of this diagram.
But, briefly, we review the familiar parts as well.
Documentation linters (such as HateDetector~\cite{winchester2023hate}) operate in the same computational manner (over text) as traditional linters do, but abstract away from them by considering text.
McNutt \etals{}~\cite{mcnutt2020surfacing} statistically mediated metaphor breaks from the purist evaluation perspective by allowing analysis to include execution.
FarSight~\cite{wang2024farsight} intersects these perspectives to provide an evaluation (and therein warnings) of potential harm from particular prompts by recalling relevant news articles via a model.
Color Buddy~\cite{mcnutt2024mixing} offers a (input) rebellious perspective on evaluating color palettes by using analysis of colors (which have a text representation, but are something different than text) to provide feedback on potential issues that palettes might have.
Next we consider the new nonuplets in depth.

\subsubsection{Evaluation Neutral / Input Rebel}
In the video game ArtSqool~\cite{artSqool}, you play a student learning to practice art, particularly in the medium of two-dimensional art. Your efforts are periodically judged by a inscrutable (and satirical) professor whose criteria are opaque.
If the criteria are not met your submission is rejected.
Viewing the game not as linear ramp, but instead casual creativity instrument~\cite{compton2019casual}, we can see this professor as a linter---albeit a suboptimal one. It provides checks if your submission is valid, its feedback can not be customized, it does not offer reasons (outside of some high-level metrics) for its complaints, and does not offer a means for fixing it\footnote{The specific framing within the game is as \emph{art critique}, rather than evaluation, which carries with it a complex set of assumptions (which include viewing critique as a facet of the art-making process rather than an end goal). The connection between critique and linting is a natural one, however we do not explore this metaphor in this work, although future work might do so usefully.}.
While still being run on a computer (and therefore evaluation neutral by \autoref{fig:alignment-chart}'s framing) this casts the Art Sqool professor not as unassailable judge, but fallible arbiter---analogous to Clippy~\cite{whitworth2005polite}.
Breaking away from objects with text representations that are evaluated through unknown processes allows us to begin seeing linting as something much broader than its traditional spell checker-like functionality.

\subsubsection{Evaluation Rebel / Input Purist}
Closely aligned with traditional code analysis is the use of downvotes on StackOverflow to indicate low-quality questions or responses.
Typically Stack Overflow content consists of a mixture of code and text. While traditional code linters (or documentation linters for that matter) might be applied to these posts, such evaluations can not cover their synthesis, or perhaps, more importantly, their context---\eg{} is an answer good if it is a duplicate of three other ones?
Fortunately, the other users of StackOverflow can offer a form of human-powered evaluation via downvotes.
This type of response is checkable---it runs and can be seen---it has a loose notion of blame---as the down vote is attached to a specific post rather than the whole page (although the specific issue within a post or comment has to be called out textually rather than through UI elements).
This form of feedback does not offer automated fixes (although sometimes follow-up comments will include specific ideas) but can be customizable through dialog with the other commenters (perhaps offering a justification on why your identical answer is better than the others).
This example highlights the connection between linters and moderation: a policy is being applied, but the context in which it is being used is important.

\subsubsection{Evaluation Rebel / Input Rebel}
Planning review boards for architectural plans offer a similar human-powered execution of a predefined standard.
Within many municipalities there exists a building code: a collection of laws saying what can and can not be built. For instance covering the maximum number of stories a building can have, how much clearance is needed for hallways, the number of required elevators for a given space, and so on.
Typically, a developer or architect is required to take their plans for a new building to a review board which considers whether or not the plans are in adherence with the building code (checkable) and highlights which parts are out of compliance (blamable).
Interestingly, this type of feedback is customizable, as the planning boards will sometimes offer variances for breaking particular laws---for instance allowing a new tower to exceed the building envelope (sometimes in exchange for a proposed civic good, such as donating funds to cap a noisy highway).
This human-powered approach to compliance does suggest that automation may be possible (such as via linter for AutoCAD), but the complexity of applying and navigating these building codes is an impediment necessitating human computation.

\subsubsection{Example cases} Before we get to the main topic, let's consider if a couple of everyday things are linters.

First, are the police a form of linter? Clearly they can evaluate behavior (such as speeding) and can provide specific blamable explanations on why they issue a ticket (such as going 100 MPH in a school zone). They can not be customized (unless you view activities like giving that arresting officer a hefty bribe to have not seen you in that school zone as a customization) and they can not provide automated fixes (they only judge past behavior which can not be altered (excluding the likes of 1994's Timecop~\cite{timecop})).
More broadly, unlike algorithms, street-level bureaucrats (\eg{} cops) refine their decision-making before making a decision \cite{alkhatib_street-level_2019}.
Linters are precisely the opposite of this inconsistently applied, uncorrectable feedback.
We suggest then that the police are not linters.

Another example: at the top of many papers on arxiv (such as on the preprint for \citet{lee2023only}) is a badge that warns, ``\emph{Important: e-prints posted on arXiv are not peer-reviewed by arXiv; they should not be relied upon without context to guide clinical practice or health-related behavior and should not be reported in news media as established information without consulting multiple experts in the field.}''
Is this a lint? It certainly is a check applied, and it can be easily ignored (for instance as it is irrelevant to the aforementioned paper~\cite{lee2023only}), although there is no remedy (automated or otherwise) or blame (what part of this paper yielded this warning?).
It is clear that all lints are warnings of some kind: a failing lint indicates that a policy, law, or standard \emph{may} be broken; not a declaration of a failure but an indication that something may be wrong.
But are all warnings lints? Again, we return to the idea that lints offer feedback on something that can be modified rather than only past behavior.
So for instance, from an author's perspective, there is nothing we can do that article to get arxiv to remove that badge (short of taking it down, which is not the expressed outcome of the warning badge). From a reader's perspective, it might be taken as a lint to their reading of the work, much in the same way that we consider for Community Notes.

Finally, is Clippy a linter?
Clippy was a widely unpopular anthropomorphized paper clip that tried to assist in the use of Microsoft Docs.
In his retrospective on why Clippy broadly failed, \citet{whitworth2005polite} argued that Clippy sealed his  fate by being impolite. He did not respect your agency, did not remember or honor your choices, did not offer useful feedback, and so on.
We see linters as being a form of polite technology: your choices to ignore them are maintained (\eg{} consider the ``add to dictionary'' button in spell checkers), they offer commentary salient to the domain (and not another), and moreover your agency is centered compared to theirs. There is, of course, an automation bias to consider (it is not polite to be domineering~\cite{mcnutt2020divining}).
While there has been some work in this direction~\cite{Correll2021ClippyDesign, heer2019agency},  developing a sociotechnical theory of respect that encompasses those factors is out of the scope of this work.

\begin{figure*}[t]
  \centering
  \includegraphics[width=\linewidth]{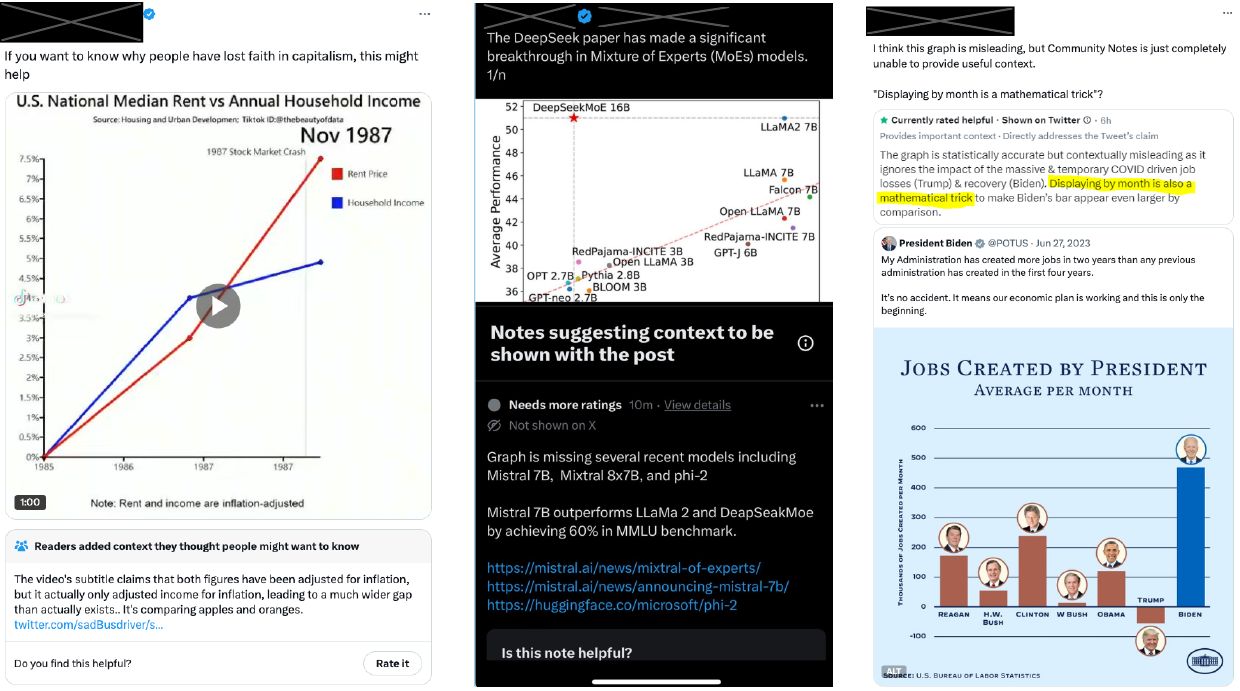}
  \caption{Example of three Community Notes as applied to data visualizations. }
  \Description{Three screenshots of tweets arranged in a row. Each tweet includes a visualization, a community note about that visualization, and a the content of the post itself. Each of them have been anonymized slightly.}
  \label{fig:community-note-example}
\end{figure*}

\section{Exploration: Community Notes as Linters}\label{sec:community_notes}

Now we come to the main topic of this work, Community Notes as linters. While community linters are conceivably usable as a linter for any domain---such as truth in general---we focus on Community Notes evaluating data visualizations. In this section, we'll briefly describe what a Community Note (formerly Birdwatch\footnote{\url{https://blog.x.com/en_us/topics/product/2021/introducing-birdwatch-a-community-based-approach-to-misinformation}}) is and how it is materialized on a social media platform. Then, we discuss how Community Notes operate as linters on data visualizations. In Figure~\ref{fig:community-note-example} we provide three examples of Community Notes that we will argue act as linters on data visualizations.

\subsection{Anatomy of a Community Note}\label{sec:community_desc}

Here, we provide a brief overview of Community Notes that as relevant to our inquiry. The idea of Community Notes can exist on any social platform (and may become more prevalent in the future), but, at the time of this writing, the X platform has the most widely used and visible implementation.
As such, we rely on their particular approach in this paper. A more nuanced description of Community Notes and the details of their creation, approval, and distribution can be found online\footnote{\url{https://communitynotes.x.com/guide/en/about/introduction}}.  Importantly, Community Notes, in their present form, only seek to add contextual information. They do not actively moderate content, for example, by removing or modifying it---this is something only that the original content creator can do.

\subsubsection{Generation \& Distribution.}
Any individual can request a Community Note be generated for some content, but at the moment, only a verified collection of individuals (Community Notes contributors) can write notes.
Community Notes are written only if there are sufficient requests for a community note to appear next to some content (\eg{} one person flagging problematic content is not enough to trigger notage).
When a note is written it is then rated by community members as ``helpful'' (Options: Yes, Somewhat) or ``not helpful''. Identifying truly helpful and unhelpful Community Notes is not just a matter of aggregating votes, applying a threshold, and filtering for helpful notes. Instead, the data gathered from all contributors and generated notes are modeled to produce a final score of helpfulness.

We do not dive into the modeling details, but it is an interesting read on how to manage sparse data while also taking into account individual preferences and variances from community contributions, without explicitly asking for them.

Whether a helpful Community Note is displayed is still dependent on a variety of factors beyond simple `majority rules'. These factors can be different across platforms. In the current X implementation, Community Notes are also evaluated based on a user's prior rating of notes. The idea is that individuals whose prior ratings are highly correlated likely have similar perspectives compared to individuals with more divergent ratings. The recency of content also plays an important role in determining the utility of Community Notes. All of these factors produce a ranking that is ultimately employed to determine whether to display a Community Note along with some content.  When a Community Note is displayed it is featured prominently with the original posted content (as in \autoref{fig:community-note-example}).

\subsubsection{Comparison to other Social Media Features.}
Community notes are not the only means that people can use to provide feedback on some content (or data visualizations). It is possible to post a response to the original content in two ways.
`Replies' can provide the critique in a thread that follows from (and links to) the original content.
Alternatively `Quote Posts' will embed the original content with a person having the option of posing a critique.
Quote Posts can act similarly to a Community Note, in that the critique is placed prominently alongside the content, although it lacks the stamp of officiality that Notes have.
Both replies and quotes may contain information that could constitute linting, but, it is not clear to what extent this is considered an effective example of human computation.

Like Community Notes, `replies' and `quotes' can be posed by anyone, but, unlike Community Notes, they are not directly vetted by a broad group. Indirect methods could substitute the `wisdom of the crowd', for example, by `liking' or `reposting' content, which may cause  a critique to gain traction.  Thus, in some respect, `replies', and perhaps more pertinently `quotes', can also function as a way to leverage human computation for linting activities. But because much of this is indirect as a side-effect of algorithmic content curation, rather than a direct and more focused community effort, we do not consider them further.

\subsection{Human Computation and Linting}\label{sec:community_linting}

Having established that Community Notes are indeed a platform for human computation, by directly recruiting from a broad community and achieving a consensus result (the content and presence of the note), we now return to a central question of this paper: \textit{are Community Notes linters?}

\subsubsection{Examples of Sociotechnical Lints.}
The three Community Notes in Figure~\ref{fig:community-note-example} broadly present what we would consider to be three examples of lints.
1) In the US National Median Rent vs Annual Household Income animated data visualization, the Community Notes indicate that the two lines have had different statistical treatments. The red line, representing rent price has \textit{not} been adjusted for inflation while that household income \textit{has}, thus exaggerating the differences. The magnitude of the difference is essentially a mirage that results from manipulated scales~\cite{mcnutt2020surfacing}.
2) Shows the performance of different large language models. The Community Note identifies several missing models that, at the time, outperform some of the top models presented in this visualization (an example of Cherry Picking).
3) The data visualization shows the rate of job growth, per month under different administrations. The Community Note indicates that this choice of aggregate measure is potentially misleading and does not account for other contextual factors (\eg{} the volatility of the pandemic). What is perhaps equally interesting in this example is the use of other social features (a quote response) to critique the ambiguity of the Community Note.

\subsubsection{But is it linting?}
Each of the examples shown \autoref{fig:community-note-example} points to legitimate issues with the data visualizations that, if valid (more on this in a moment), provide important information for the creators and readers of these charts.
These examples concern scale, missing data, and choice of aggregation have been explored in prior research for linting visualizations~\cite{mcnutt2020surfacing, chen2021vizlinter} and machine learning datasets~\cite{hynes2017data}.
However, in both instances, purely automated strategies (\eg{} metamorphic testing, a deep neural network, static analysis) are proposed. Here, the same kinds of evaluations (or lints!) emerge organically in Community Notes.

However, we can go further. In \autoref{fig:alignment-chart} we categorize Community Notes as fitting in the Input Neutral and Evaluation Rebel alignment as their input is (mostly) text and their means of evaluation uses humans (organized by the community notes algorithm). We can also now show that Community Notes conform to the expected behaviors of linters, as outlined in the linting ladder:

\begin{description}
\item[\textit{Community Notes are \textbf{Checkable}.}]
        Notes check against social norms in general, but specifically against misinformation.
        Interestingly, Community Notes provide no prescriptive definition of \textit{what} misinformation is.
        The decision is made by the community (echoing our notion of linters as community standards).
        For data visualizations, it is entirely possible for people to also provide lints on adherence to visualization design principles or guidelines, among other previously established forms of visualization linters~\cite{mcnutt2023design, chen2021vizlinter} as well.
        But these are almost secondary considerations to the larger question of whether the visualization is misleading.

\item[\textit{Community Notes are \textbf{Customizable}.}]
        The presence of the note next to some social media content is dependent on the community evaluation, which is in turn established by several criteria.
        While (currently) individuals using a social media platform do not have the option to selectively deactivate lints, the community does.
        For example, any individual is free to propose a Community Note, but if there is not enough broad support, the note never materializes  -- in other words, the lint is deactivated. The ranking algorithm can also be adjusted to further customize whether a Community Note is shown or not. It may be an interesting thought experiment to consider the implications of deactivating social links at the level of individual users (such as via an ad block-style browser extension). Can (should?) a user request to ignore a Community Note in their timeline? Or will this merely filter into a linty bubble?

\item[\textit{Community Notes are \textbf{Blamable}.}]
        Community notes can provide specific examples of the issue switch a data visualization, as is shown clearly in Figure~\ref{fig:community-note-example}. While the relative precision of these lints is not specified by the format, these can go into details specifics, or highlight generality.

\item[\textit{Community Notes can Support \textbf{Fixable} Changes.}]
        In evaluating this rung, we debated to what extent Community Notes provide fixable suggestions. One mechanism for fixing misleading content is through social shaming, whereby the original user will actually delete the original content that a Community Note identified as misleading. However, this represents an extreme change -- a standard linter would not suggest you should delete your entire code, but suggest more nuance refinements. Editing capabilities are inconsistent on social platforms, but, when they exist, they could provide a more nuanced mechanism to fix original content.
\end{description}

It would seem then that we can see community notes as linters because they do lint-like things (\ala{} being a good person because you do good things, rather than doing good things because you are good person).
More broadly they exist with a comparable social contract, as they embodying a dynamic framework for social accountability and refinement.
They are checkable, aligning with social norms to identify misinformation, customizable, adapting to collective evaluations, blamable, by pointing out specific shortcomings, and and can even guide fixable improvements.
As seen here, community-driven processes can be used to enhance content quality---whether for social media or visualizations or whatever ever---by balancing the rigor of shared standards with the flexibility needed for iterative progress.
This interplay of community-driven critique and iterative improvement underscores the broader potential for collaborative systems to foster trust and transparency in diverse domains.

\subsection{Beyond Traditional Linters}
\label{sec:impliciations}

If we accept that Community Notes are indeed a form of linting powered by human computation, then, they also present some interesting implications
beyond what is considered with traditional linters.

A prominent potential bias in linters is the assumption that if the linter passes, there are no problems with the object of analysis.  Community-supported linters exhibit an interesting variation on this: if one post sports a Community Note, does that imply that all other posts that do not have Community Notes are fine? Does this broadly imply that everything else in the universe has been deemed acceptable? While this seems absurd, this is more specifically known as the implied truth effect~\cite{pennycook2020implied}. In the Community Notes algorithm, the factors relating to whether a note appears or not are related to contextual factors (\eg{} recency, engagement) that fail to prioritize misleading content even with the same type of lint (\eg{} visualizations with scale issues are not consistently flagged). Those present an interesting opportunity to explore a hybrid model, where, once humans identify relevant lints an automated system backfills the rest. Such a mechanism presents its own challenges but is an interesting avenue to explore.

Another interesting difference is how linters respond to unresolveable issues. Something we also observed within our work on linting color palettes (as well as navigating some troublesome ESLint configs), is that in some cases collections of lints are not always resolvable. This can cause the user to fix one issue, and then, in attempting to fix another issue, break the first. An example of an attempt to resolve this issue is VizLinter~\cite{chen2021vizlinter}, which tries to propose fixes that fix not just one linted issue by all subsequent ones that arise from fixing that initial issue---however this does lead to strong automation bias.  This type of conflicting domain knowledge appears in the context of a Community Note as well. For instance, one could view mutually conflicting communal opinions on a given post as being an example of this.  The current approach taken in the Community Notes algorithm is to look for consensus among divergent community bases (\eg{} people with different political affiliations both agree that some content is problematic). However, this is an interesting social choice and one that hides the complexity and implications of what may be unresolvable lints. For example, there may be two sizable camps, the first believes pie charts should never be used and another that believes that they are a broadly approachable graphical form. Such conflicts and differing community standards are technically modeled but not presented with a Community Note, and is interesting future work.

Finally there is a question of \emph{who} is doing the linting, which raises questions both of expertise as well as adversarial behavior.
In traditional linters the collection of lints are typically prepared by someone expert in that domain.
For instance, \citet{lei2023geolinter} describe a protracted engagement with a geographer to design their lints for maps.
By their nature, community notes draw on individuals who are not necessarily expert in all areas.
\citet{kangur2024checks} describe how external sources are used as validation (to various effects).
This situation is more complicated within visualization, which lacks unambiguous means of evaluating visualization quality (either automatic or theoretical). While guidelines exist~\cite{visguides} their validity is often debated~\cite{correll2014bad}.
Similarly, the potential for misinformation inherent to this setting creates opportunities for antagonistic actors to create misinformation through the notes themselves. The communal aspects of community notes attempt to circumvent this issue by drawing on the amiability of the crowd.
Potentially these issues suggest that something akin to means of external sourcing for visualization analysis are necessary to form useful commentaries.

\section{Discussion}

This work poses a provocation: what happens if we understand the granular, specific, and actionable feedback typically enshrouded in linters as a computational tool, as something that can be wielded by the crowd, as a representation of their beliefs and values?
That is, one of the goals of this paper (particularly as an alt.chi paper) is to explore the effect of recontextualizing linters into a social computing area.
In doing so we sought to explore this metaphor in closer depth and more broadly understand its bounds.

Generally we find this metaphorical extension to be informative: it usefully reveals some new things about linters (such as the requirement that they operate over mutable things) and highlights the similarity with tools that we find in more everyday circumstances.
Following Carpenter~\cite{they_live}, we hope that in posing this thought experiment we prompt the reader to see the world in a new light: what things existing in ambient day-to-day life are linters?
Of those that are, what can linters learn from them, and what can they learn from linters (such as  fitting onto the rungs of the linting ladder)?
Of those that aren't, would making them more lint like be useful?
We close this paper with several considerations.

\subsubsection{Mandatory AI subsubsection}
A recent line of work has explored replicating algorithm's using crowd-work or distributed human computation through LLMs.
\citet{wu2023llms} develop a variety of LLM-based pipelines for replicating crowd-work algorithms.
\citet{grunde2023designing} develop a broader theory of how these types of pipes parallel one another.
In parallel, AI-ification of linters is something that naturally falls out of this comparison.
Conceivably, such a combination could usefully operationalize domain knowledge without the expense of human attention.
Moreover, it is not without precedence: the original work on AI Chains~\cite{wu2022ai} explored a critique system for visualization that is viewable as a linter.
\citet{jupybara25Wang} employ a collection of lint-like critics to guide the automated creation of visualization narratives.
Human-powered linters seem best suited to problems that are generally difficult to automate or scale, such as moderation~\cite{gorwa2020algorithmic}.
In embracing new technologies, such as AI, to enlarge the remit of linters, we should be careful to consider if such methods are being used to actually expand linters' grasp or merely simplify their implementation.

\subsubsection{Linters and uncertainty}
As with any human driven computation, there will naturally be some variance compared to the diligent application of programmatic logic found in computer-driven computation. While the algorithmic implementation of tools like Community Notes reduce this variance, this may also give way to implied truth effect biases.  While socially mediated tools like Community Notes communicate this uncertainty through a connection with a human element (people are only human and make mistakes), other areas in which human computation might be applied as linter might not do so as clearly. It seems important then, in developing future human-powered linters  to consider means of communicating this uncertainty and not let faux-automation (or just dogmatic presentation) lead to unintended ends.
Beyond the human linters, communicating uncertainty in lints is a useful future research for lint presentation (although \citet{hopkins2020visualint} explore a related visual metaphor with sketchy rendering) as some tools may give back variable responses---such as in using statistically motivated evaluation methods, as \citet{mcnutt2020surfacing} do, or in using non-deterministic AI-based methods.

\subsubsection{Linting Philosophy}
Just as all scientific works have a notion of truth that motivates their work, so too do assistant-style tools carry epistemologies of different sorts.
For instance,  recommendation systems position themselves in such a way a that suggests that there is a correct answer among a possibility space. In many ways, linters are the dual of recommendation systems: rather than suggesting they evaluate, they approach correctness in a manner which is undogmatic and can be overridden following the user's desires (\cf{} \autoref{fig:lint-euler}).
\citet{offenhuber2023autographic} describes a space of visualizations that form themselves---for instance desire footpaths being a (autographic) visualization of people's preference for cutting corners and so forth. Linter's evaluative perspective is hard to reconcile with this naturally occurring process perspective, and so additional consideration of the epistemological underpinnings of linters is needed in order to form comparable evaluations. This may not be generally reconcilable, as linters assume that ``incorrect'' designs may be fixed and adjusted, which is not necessarily the case for such systems (what would an incorrect desire path mean?).

\subsubsection{Limitations and future work}
This work offers a reflection on a specific form of linters---from a visualization rather than social computing perspective. While this serves our goal of offering a provocation, it does not explore how general users understand content presented in that way.
While there are quite a number of studies about Community Notes (\eg{}~\cite{kangur2024checks, drolsbach2024community}) and their relationship with trust, ours is the first to investigate this alternative viewpoint.
Understanding the role of trust in visualization~\cite{elhamdadi2023vistrust} is a pressing issue, and further consideration of the connection between trust, human-powered moderation-like tools, and visualization seems like useful future research.
One means of increasing the trust in visualizations presented on social media might be to create a new social-media platform in which only visualization researchers are allowed to post (perhaps VizSky, VIStter, or V). This will ensure that the graphics presented are of high quality, but in a manner that may be  a little boring.

\begin{acks}
  We are grateful for the pointers provided by Maxim Lisnic.
  We appreciate the interesting commentary from our many reviewers.
\end{acks}

\bibliographystyle{ACM-Reference-Format}
\bibliography{./community-lints}

\end{document}